# NORMAL-FAULT STRESS AND DISPLACEMENT THROUGH FINITE-ELEMENT ANALYSIS

A. Megna[1], S. Barba[1], and S. Santini[2]

[1] Istituto Nazionale di Geofisica e Vulcanologia, via di Vigna Murata 605, Roma, Italy

[2] Università degli Studi di Urbino, Istituto di Fisica, via S. Chiara 27, Urbino, Italy

## *Abstract*

We compute displacement and stress due to a normal fault by means of two-dimensional plane-strain finite-element analysis. To do so, we apply a system of forces to the fault nodes and develop an iterative algorithm serving to determine the force magnitudes for any slip distribution. As a sample case, we compute the force magnitudes assuming uniform slip on a 10-km two-dimensional normal fault. The numerical model generates displacement and stress fields that compare well to the analytical solution. In fact, we find little difference in displacements (<5%), displacement orientation (<15°), and stress components (<35%, half of which due to slip tolerance). We analyze such misfit, and discuss how the error propagates from displacement to stress. Our scheme provides a convenient way to use the finite-elements direct method in a trial-and-error procedure to reproduce any smooth slip distribution.

## *Introduction*

The displacement, deformation, and stress fields due to earthquakes can derive from analytical or numerical methods. The analytical methods differ among each other mainly because they model different properties of the medium or use different numerical procedures. In the literature, we find the medium represented as homogeneous and elastic (Okada, 1992), viscoelastic (Dragoni *et al*., 1986; Pollitz, 1997), or layered (Morelli *et al*., 1987; Bonafede *et al*., 2002), whereas the numerical procedures include numerical integration (Sato and Matsu'ura, 1973) or finite series of Lipschitz–Hankel integrals in a recursive algorithm (Ben-Menahem and Gillon, 1970; Rybicki, 1971). To reduce the calculus complexity, semi-analytical methods face problems with few layers, where the dislocation lies entirely within one layer or within the half-space (Ma and Kusznir, 1995; Savage, 1998). For cracks and strike-slip faults, Bonafede *et al*. (2002) solved the problem of a dislocation cutting the interface between one layer and the half-space. On the other hand, numerical Green's functions help solving problems where arbitrary faults cut a multilayered medium (Wang *et al*., 2003). All the above works deal with horizontal layers.

To include lateral heterogeneities into the model, the finite element method (FEM) is a better tool (Masterlark and Wang, 2002; Bielak *et al*., 2003; Bustin *et al*., 2004). The FEM allows dealing with layers over a half-space (Cattin *et al.,* 1999), with inhomogeneous crustal structures (Barba, 1999; Carminati *et al*., 2001; Zhao *et al*., 2004), and with physiogeographic characteristics (Tinti and Armigliato, 2002;

Armigliato and Tinti, 2003). In the case of uniform slip along the fault, a few authors (e.g., Cattin *et al.,* 1999, and Zhao *et al*., 2004) compared results coming from FEM with those derived by analytical methods. These works mainly focus on surface displacements, but leave out displacement and stress fields at depth. As a variation, we develop a procedure to constrain any slip distribution by applying force couples of different magnitude at the sides of the fault and reaction forces orthogonal to the fault. To compare our results with the Okada's analytical solutions, we build a two-dimensional plane-strain finite-element model and, assuming the simple case of uniform slip, we determine the magnitude of the required forces. We represent displacement and stress fields at the free surface and in the vertical section across the fault. In this paper, we intend to test our technique in simple cases, before applying it to further and more complex seismological problems.

### *Design of the Finite Elements Model*

To allow one-to-one comparison with analytical methods, we model the dislocation of a normal fault in a homogeneous, isotropic, and elastic medium. We build the two-dimensional finite-element model in a vertical section across the fault and compute, assuming plane strain, the displacement and stress fields by using the program MSC.Marc® (MSC.Software, 2003). MSC.Marc solves the equation of motion for continuous bodies, yielding a Lagrangian description and the Newton-Raphson iterative method. As a typical normal fault, we set the following parameters: 10 km width, 40° dip angle, 3 km top depth, 1 m uniform slip.

Comparing a two-dimensional numerical model with the analytical three-dimensional case requires evaluating the role of the fault length. To this purpose, we choose three different lengths: 10 km, 28 km, and 500 km: 10 km represents a typical

M~6.0 earthquake fault that can give surface effects observable in the field (e.g., in Barba and Basili, 2000), whereas 28 km allows comparing our results with those of Cattin *et al.* (1999). On the other hand, the 500 km length helps avoiding edge effects in the central section, where we can approximate the model as two-dimensional. To compare with the analytical half-space, the finite-element model must be much larger than the fault. Assuming a fault width of 10 km, we set the model width to 300 km and its depth to 100 km. As of model boundaries, we impose zero orthogonal displacement at the bottom and lateral edges (Figure 1a), whereas the free surface complies a stress-free condition. Typical elastic properties sketch the rheology of the crust (Poisson ratio $\rho=0.27$; Young modulus $Y=10^{11}$ Pa). We mesh the model through four-node quadrilaterals (4290 elements and 4433 nodes), refine the grid next to the fault, and coarsen it otherwise (Figure 1b). The smallest elements (0.2 km) lie along and near the fault, and between the fault and the free surface. These choices allow us to achieve enough accuracy in the solution and to limit the error propagation.

The origin of the Cartesian coordinate system lies on the free surface above the fault upper tip. The section lies in the y-z plane, with coordinates increasing to the right and up (Figure 1a) - the medium lies in the $z\leq0$ domain. The fault cuts the mesh, providing a free-slip interface embedded into the crust. We obtain the free-slip interface by (1) duplicating the nodes on each side of the fault and not allowing the two fault edges to intersect, as in Melosh and Raefsky (1981), (2) assuming zero friction at the interface, and (3) forcing duplicate nodes to stay on the fault line. Inner normal forces appear and tend to separate the two fault edges: we impose that, on each node, a reaction force balances the separation force to satisfy condition no. 3. We set an upper limit ($10^{12}$ N) to the reaction force that is never met. Therefore, the fault does not break in tension. Because of the zero-friction condition, reaction forces are orthogonal to the fault sides and oriented towards the fault. We do not set any constrain on the fault dip or position, so expecting a non-uniform dip to result in the final iteration. We apply the two forces of each couple at the opposing nodes of the fault (Figure 2a), assuming direction parallel to the fault but opposite verse on each side. The magnitude instead depends on the position, on the material properties, and on the slip distribution that we impose. The magnitude of the reaction forces depends on the position too. The system of slip-parallel and reaction forces plays the role of the classical "double couple" acting on the fault – with the exception of the fault tips. The double-couple total moment tends to zero when the distance among slipping nodes tends to zero, i.e. for very small elements. In practice, "finite" elements have a finite size, but we verify that the small resulting moment does not cause unsought effects.

In order to get any slip distribution we include the following algorithm (thanks to the user subroutines option) into the MSC.Marc program, where we formulate a static problem and use the "time" as an iteration index. Let's define the wanted slip distribution as $U_n$, the node index as n (n=1,…,N), the average slip as $<U_n>$, and the iteration index as t (t=1,…,T). Moreover, let's use + and – (plus and minus) to indicate respectively nodes at the right and at the left of the fault. Therefore, $U_{nt}$ indicates the absolute slip at any node and iteration, derived from the fault-parallel relative slip between opposing nodes: $U_{nt}=U_{nt}^{+}+U_{nt}^{-}$. In the first iteration, we apply force couples having the same magnitude everywhere on the fault. To make the computation stable, we set the initial magnitude in order to get $<U_{nt}> \leq <U_n>$. With respect to the uniform slip distribution $U_{nt}=<U_n>$, at the beginning the slip in the central part is larger than near the tip of the fault. Subsequent iterations allow varying, in a trial-and-error procedure, the magnitude of the forces, until $<U_{nt}>-<U_n>$ decreases under a fixed tolerance ($<U_n>/100$). The resulting force magnitude increases from the centre to the tips (Figure 2b). In our model, where we require $<U_n>=1$ m, the force ranges from $2\cdot10^9$ N to $10^{11}$ N at the tip. We find actual slip tolerances of 0.1% – 0.9%, with greater values at the tip of the fault, whereas slip at adjacent nodes differs ~1 mm in the average. Therefore, our algorithm works for any smooth slip distribution.

## *Misfit between analytical and numerical solutions*

For a uniform slip distribution, we compute displacement and stress in a two-dimensional section across the fault. In the analytical computation, the fault has a finite length (L = 10 km, 28 km, and 500 km) and the section cuts the fault in the middle. Concerning the vertical displacement at the surface, the coseismic uplift increases with the length of the fault, whereas the subsidence does not follow a simple rule (Figure 3a). The numerical solution compares well with the L=500 km analytical one but, in the uplifted area, the numerical solution goes to zero faster as the distance from the fault increases. We ascribe the differences between the numerical and the analytical solution mostly to inaccurate meshing between the fault tip and the surface, as we find high misfit where absolute values are small. On the other hand, the horizontal displacement derived through the numerical solution reproduce very well the analytical solution (L = 500 km). The displacement increases with the length of the fault, especially at distances of ~10 km and more (Figure 3b), where the length dominates the result.

To evaluate the accuracy of the "numerical" displacement field, we refer to the analytical solution (L = 500 km), determining the misfit of the displacement vector magnitude and orientation for each node in the mesh. The displacement decreases with the distance from the fault, with most of it occurring in the hanging wall (Figure 4a). This asymmetry depends on the free surface. The presence of the fault, acting as a discontinuity in the displacement orientation, makes the accuracy hard to evaluate in the immediate surroundings of the fault – a small error in position can show as a large error in the misfit computation. Therefore, we do not compute the misfit at the fault. Most of the nodes show a small displacement misfit (less than 0.02 m), with the maximum misfit (0.06 m) located in the immediate surroundings of the fault (Figure 4b). In percentage, we get less than 5% near the fault and in most of the computed values. Higher values (greater than 10%) lie between the upper tip of the fault and the free surface. We ascribe the difference, as above, to the inaccurate meshing – combined with the small displacement, which cause the model to be ill conditioned. On the other hand, such a problem occurs in a rather small region, where the absolute misfit happens to be small (less than 0.03 m) and the horizontal component of displacement oscillates near zero. Therefore, when small values are involved, we have to develop a more accurate mesh, use eight-node quadrilaterals, or prefer the analytical solution. Concerning the orientation, we define the misfit as the minor angle between the analytic and the finite-element displacement vector. We can neglect the small orientation misfit (<2°) that we find close to the fault, or the slightly larger one (typically less than 10°) that we find in most of the model (Figure 4c). But, as expected, we find a greater misfit (>30°) close to the free surface (y~4 km, z~0 km), where the displacement is small and the horizontal displacement oscillates near zero – a small change in horizontal displacement gives a large (45°) error in orientation. The lack of constraint on the fault dip and position allows our fault to rotate counterclockwise by ~ $6 \cdot 10^{-5}$ rad, near as much as the analytic solution (e.g.,

Bonafede and Neri, 2000; Armigliato *et al.*, 2003a; Armigliato *et al.*, 2003b) but much less than the orientation misfit.

As for the stress field, we compute the three components of the tensor ($\sigma_{yy}$, $\sigma_{zz}$, and $\sigma_{yz}$) and make the difference with the analytical solution. Instead of all components, which exhibit similar patterns, we discuss the misfit of $\sigma_{yy}$ only. The stress reaches the highest values at the fault tips (>$10^7$ Pa) and stays low elsewhere (Figure 5a-c). Our results compare favourably with the analytical solution: we find a small difference (<$10^5$ Pa) in most of the model, whereas a larger misfit ($10^6$ - $10^7$ Pa, or 10%-100%) occurs at the tip of the fault (Figure 5d). Here, the mesh fails to reproduce the high stress gradient because of the low nodal density: the slip tolerance alone accounts for $5 \cdot 10^5$ - $5 \cdot 10^6$ Pa misfit in stress components, i.e. 5% - 50%. On the other hand, the percentage misfit shows that relatively high misfit (>35%) occurs only where $\sigma_{yy} \sim 0$, indicating minor computation instabilities. To visualize the misfit of all stress components, we use the equivalent Von Mises stress $\sigma_{VM}$, which gives an equivalent scalar measure of the deviatoric stress tensor $S_{ij}$, $\sigma_{VM} = \sqrt{\frac{3}{2} \sum_{ij} S_{ij} S_{ij}}$, where $S_{ij}$ depends on stress $\sigma_{ij}$ as $S_{ij} = \sigma_{ij} - \frac{1}{3} \sum_k \delta_{ij} \sigma_{kk}$. In most of the model, the difference between the numerical and analytical $\sigma_{VM}$ is less than 35%, suggesting the numerical result to be stable.

## *Conclusions*

We propose here an iterative algorithm that can reproduce any smooth static slip distribution on a fault in a finite-element scheme. The procedure uses variable magnitude forces parallel to the fault and reaction forces orthogonal to it. This system of forces, whose total moment is near zero, acts as the classical "double couple".

To test our procedure in a simple case, we consider static displacement due to uniform slip across a normal fault. By means of two-dimensional finite element modelling, we compute the displacement and the stress fields. We apply a series of force couples of variable magnitude to the dislocating nodes and reaction forces orthogonal to the fault: uniform slip requires larger forces to develop at the fault tips and lesser at the centre. As a new addition to this well-known analytical problem, our iterative algorithm serves to determine the magnitude of inner forces in the finite element model. Such numerical model generates displacement and stress fields which compare well to those obtained through analytical solutions.

We compare the two-dimensional FEM results with the analytical solution in the case of a normal fault 500 km long. As of surface displacement, our model retrieves the horizontal component correctly, whereas the vertical component shows slightly different extreme values. The displacement field exhibits little absolute misfit, whereas the highest relative misfit occurs within low displacement areas. The stress field yields acceptable values, too, with generally low misfit everywhere but at the fault tips. We ascribe such differences in the displacement and stress fields to the mesh used during computation and, merely for the stress field, to our choice of slip tolerance. Real-world applications may require a different mesh, depending on the accuracy requested on each output quantity.

The algorithm presented here allows using the "time", t, as an iteration parameter in a pure finite-element scheme. It allows computing displacement and stress field induced by a static dislocation of a normal fault. Despite we deal only with uniform slip, the method can reproduce any smooth slip distribution and possibly any pattern of surface displacement.

## *Acknowledgments*

We are grateful to Michele Dragoni for the discussions and the support at various stages of the work. We thank Alberto Armigliato and an anonymous referee for the accurate review and useful suggestions that improved the paper. Thanks to Giuliano Milana for the discussion about the double-couple. Work partially funded by INGV-DPC-S2-UR3.1.

## *Figure captions*

**Figure 1.** Two-dimensional finite element model. (a) geometry (not on scale) and boundary conditions (circles represent no movement across the edge). (b) enlargement of the quadrilateral-element mesh relative to the dashed box shown in (a).

**Figure 2.** (a) Sketch of the force couples (F, continuous-line arrows) and the reaction forces (R, dashed arrows) applied to N nodes along the fault (line); each filled circle represents two opposing nodes on either side of the fault, falsely separated in the enlarged outline, where the gray shades represent the elements. (b) Magnitude of the force (F) versus down-dip distance, in the case of 1 m uniform slip.

**Figure 3.** (a) Vertical and (b) horizontal displacements at the free surface due to 1 m uniform slip. Continuous line: analytical solution; dashed line: two-dimensional finite-element model. The fault length varies, all other parameters kept constant.

**Figure 4.** (a) Displacement field computed through finite element model of Figure 1. (b) Difference (misfit) between the numerical (FEM) and the analytical displacement fields: Contours of absolute (color bands) and relative (white dashed line: 10%, white solid line: 5%) differences in value. (c) Orientation misfit (inner angle). The black solid line represents the fault.

**Figure 5.** (a) $\sigma_{yy}$, (b) $\sigma_{zz}$ and (c) $\sigma_{yz}$ components of the FEM stress tensor. (d) Absolute difference between the finite-element and the analytical $\sigma_{yy}$; (e) relative difference between the numerical and the analytical scalar equivalent Von Mises stress.

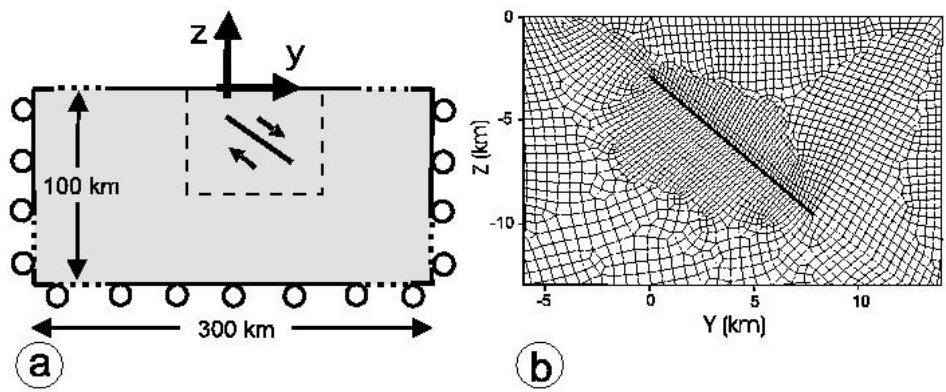

Fig. 1

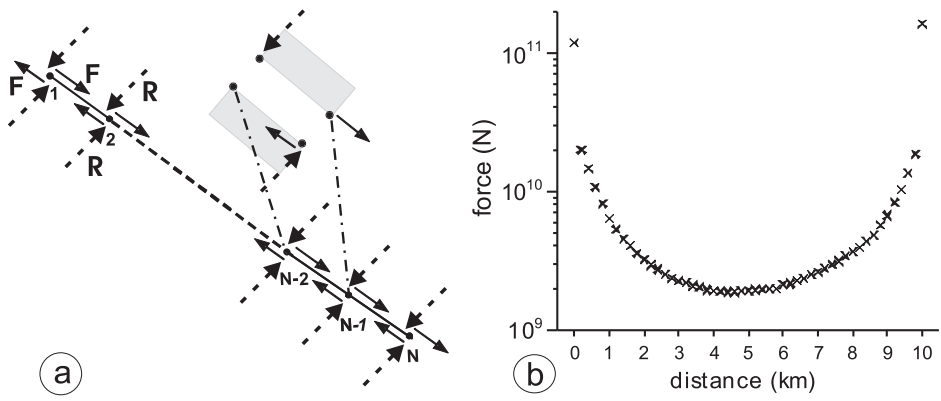

Fig. 2

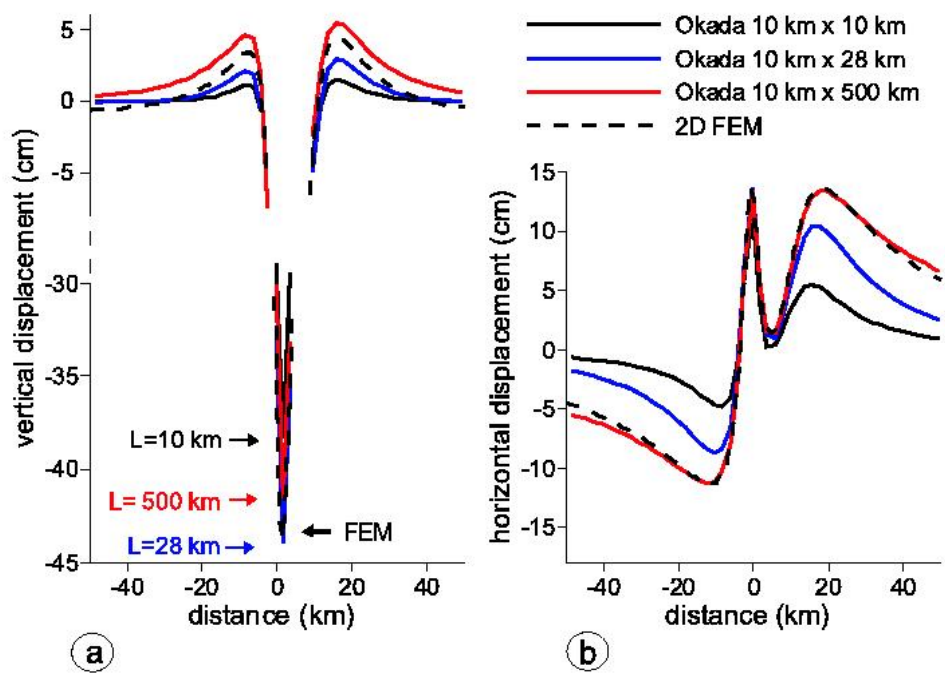

Fig. 3

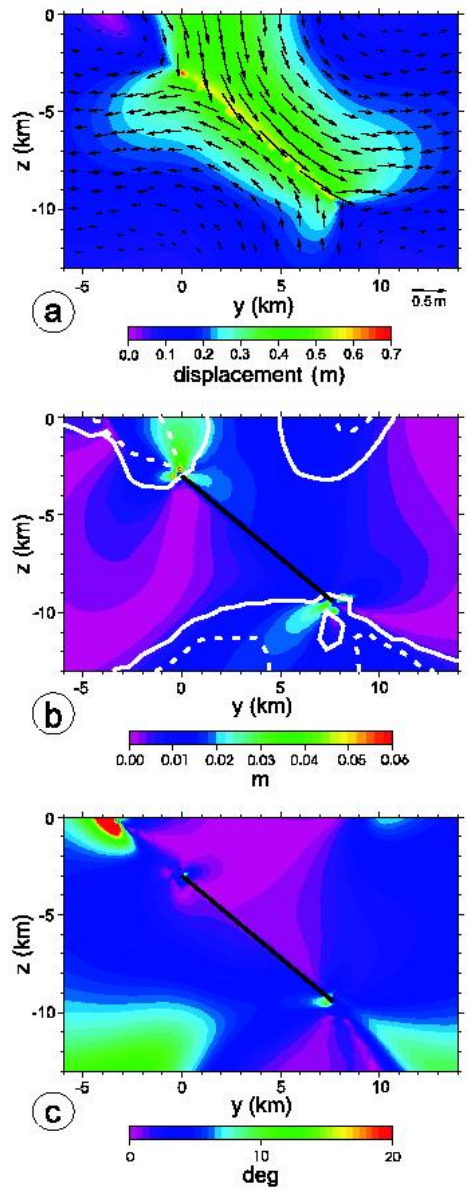

Fig. 4

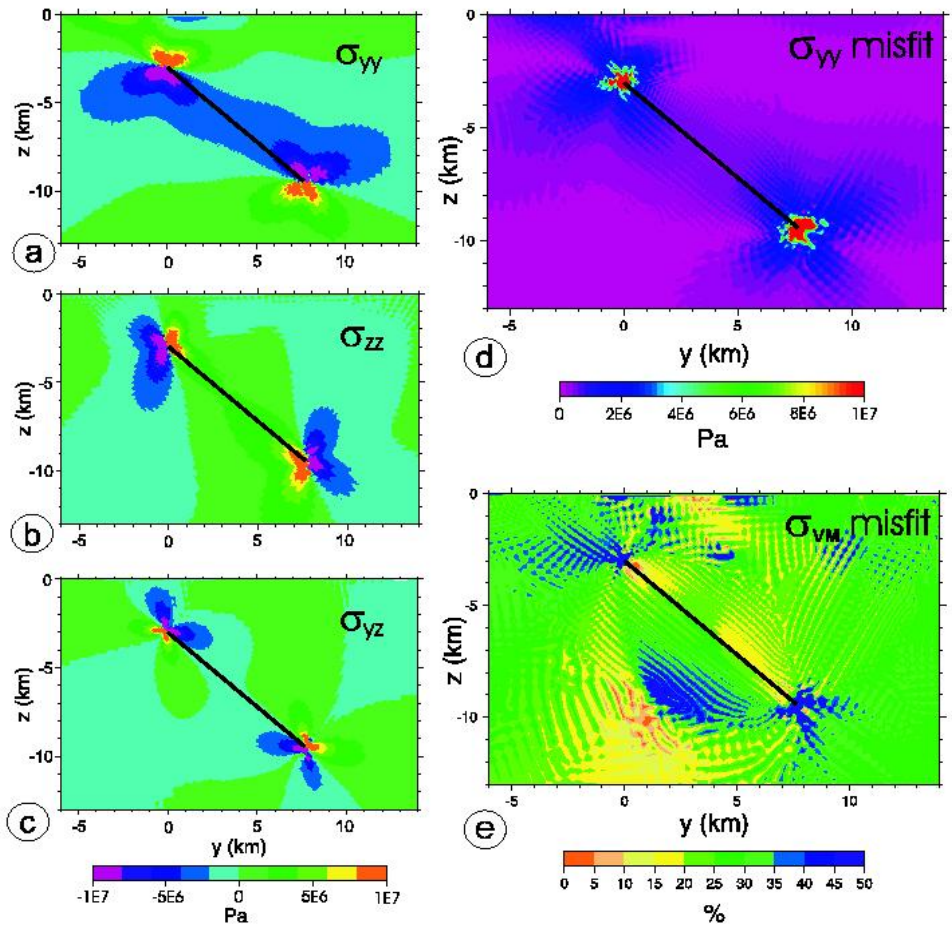

Fig. 5